\documentclass[twocolumn,superscriptaddress,aps,prb]{revtex4-1}

\usepackage{graphicx}
\usepackage[usenames,dvipsnames]{color}
\usepackage[normalem]{ulem}
\usepackage{siunitx}
\newcommand{\nit}{NIT-2Py}
\newcommand{\mub}{$\mu_\mathrm{B}$}

\begin{document}

\title{Unconventional field induced phases in a quantum magnet formed
  by free radical tetramers }

\author{Andr\'es Sa\'ul$\sp{*}$}

\affiliation{Aix-Marseille University, Centre Interdisciplinaire de
Nanoscience de Marseille-CNRS (UMR 7325), Marseille, France}
\email{saul@cinam.univ-mrs.fr}

\author{Nicolas Gauthier} 

\affiliation{D\'epartement de Physique, Universit\'e de Montr\'eal,
Montr\'eal, Canada}

\affiliation{Regroupement Qu\'eb\'ecois sur les Mat\'eriaux de
Pointe (RQMP)}

\affiliation{Laboratory for Scientific Developments and Novel
Materials, Paul Scherrer Institut, 5232 Villigen, Switzerland}

\author{Reza Moosavi Askari}

\affiliation{D\'epartement de Physique, Universit\'e de Montr\'eal,
Montr\'eal, Canada}

\affiliation{Regroupement Qu\'eb\'ecois sur les
Mat\'eriaux de Pointe (RQMP)}

\author{Michel C\^ot\'e} 

\affiliation{D\'epartement de Physique, Universit\'e de Montr\'eal,
Montr\'eal, Canada}

\affiliation{Regroupement Qu\'eb\'ecois sur les Mat\'eriaux de
Pointe (RQMP)}

\author{Thierry Maris}

\affiliation{D\'{e}partement de Chimie, Universit\'{e} de
Montr\'{e}al, Montr\'{e}al, Qu\'{e}bec, Canada}

\author{Christian Reber}

\affiliation{D\'{e}partement de Chimie, Universit\'{e} de
Montr\'{e}al, Montr\'{e}al, Qu\'{e}bec, Canada}

\affiliation{Regroupement Qu\'eb\'ecois sur les Mat\'eriaux de
Pointe (RQMP)}

\author{Anthony Lannes}

\affiliation{Laboratoire des Multimat\'eriaux et Interfaces (UMR
5615), Universit\'e Claude Bernard Lyon 1, 69622 Villeurbanne cedex,
France}

\author{Dominique Luneau}

\affiliation{Laboratoire des Multimat\'eriaux et Interfaces (UMR
5615), Universit\'e Claude Bernard Lyon 1, 69622 Villeurbanne cedex,
France}

\author{Michael Nicklas}

\affiliation{Max Planck Institute for Chemical Physics of Solids,
Dresden, Germany}

\author{Joseph M. Law}

\affiliation{Hochfeld-Magnetlabor Dresden (HLD-EMFL),
Helmholtz-Zentrum Dresden-Rossendorf, D-01314 Dresden, Germany}

\author{Elizabeth Lauren Green}

\affiliation{Hochfeld-Magnetlabor Dresden (HLD-EMFL),
Helmholtz-Zentrum Dresden-Rossendorf, D-01314 Dresden, Germany}

\author{Jochen Wosnitza}

\affiliation{Hochfeld-Magnetlabor Dresden (HLD-EMFL),
Helmholtz-Zentrum Dresden-Rossendorf, D-01314 Dresden, Germany}

\author{Andrea Daniele Bianchi}

\affiliation{D\'epartement de Physique, Universit\'e de Montr\'eal,
Montr\'eal, Canada}

\affiliation{Regroupement Qu\'eb\'ecois sur les Mat\'eriaux de
Pointe (RQMP)}

\author{Adrian Feiguin}

\affiliation{Department of Physics, Northeastern University, Boston,
Massachusetts 02115, USA}

\begin{abstract}
  We report experimental and theoretical studies on the magnetic and
  thermodynamic properties of NIT-2Py, a free radical-based organic
  magnet.  From magnetization and specific heat measurements we
  establish the temperature versus magnetic field phase diagram which
  includes two Bose-Einstein condensates (BEC) and an infrequent half
  magnetization plateau. Calculations based on density functional
  theory demonstrates that magnetically this system can be mapped to a
  quasi-two-dimensional structure of weakly coupled tetramers.
  Density matrix renormalization group calculations show the unusual
  characteristics of the BECs where the spins forming the low-field
  condensate are different than those participating in the high-field
  one.
\end{abstract}

\maketitle

\section{Introduction}

The exact mapping between spin $S=1/2$ systems and hard bosons
proposed by Matsubara and Matsuda in 1956 \cite{Matsubara1956} has
opened the possibility of observing Bose-Einstein condensates (BEC) in
quantum magnets.  Several experimental realizations can be found in
the literature, very often formed by interacting transition metal
dimers \cite{zapf_bose-einstein_2014,giamarchi_boseeinstein_2008}.  A
typical scenario invokes a ground state described by pairs of
localized spins forming singlets.  An external magnetic field acts as
an effective chemical potential for triplet excitations that can
subsequently form the BEC, characterized by the presence of (XY) long
range magnetic order in the direction perpendicular to the field.
Since a finite magnetic field $H_{c1}$ is necessary to break the
dimerized singlets, the temperature versus magnetic field phase
diagrams typically display a ``dome'' structure bounded by two
critical fields, $H_{c1} < H < H_{c2}$, and a field-dependent critical
temperature $T_c(H)$.  While most cases of magnetic BECs formed by
$S=$ 1/2 dimers follow this picture (see
Ref.~[\onlinecite{zapf_bose-einstein_2014}] for a review) there are
magnets such as Cs$_2$CuCl$_4$
\cite{Coldea2002,Radu2005,Sebastian2006}, where the system is already
ordered at zero field.
  
BECs have also been observed in systems formed by $S=$ 1 dimers such
as Ba$_3$Mn$_2$O$_8$ \cite{Samulon2008,samulon_asymmetric_2009} or the
organic biradical F$_2$PNNNO
\cite{Tsujii2005,Hosokoshi1999,Bostrem2010} where the total spin can
take the values 0, 1, and 2.
These systems present an energy gap above the singlet ground state and
a half magnetization plateau corresponding to the triplet state of the
dimers. In Ba$_3$MnO$_8$ \cite{Samulon2008,samulon_asymmetric_2009}
two field-induced domes have been observed, the first one
corresponding to the condensation of triplets and the second one to
the condensation of quintuplets.

A similar behavior with two field-induced domes can be expected in a
system of weakly interacting $S = $ 1/2 tetramers, where it is
possible to realize non-trivial intra-tetramer quantum order
determined by the relative strength of the exchange interactions.
Unfortunately, contrary to the large amount of low dimensional systems
where the magnetic centers form dimers, there are very few low
dimensional systems formed by interacting tetramers:
Cu$_2$CdB$_2$O$_6$ \cite{Hase2005,Hase2009,Hase2015,Janson2012},
CuInVO$_5$ \cite{Hase2016}, and SeCuO$_3$ \cite{Zivkovic2012}. In
these $S=1/2$ systems, where the magnetic centers are $d$ electrons
carried by the Cu atoms, the large values of the magnetic interactions
prevents the experimental exploration of the full phase diagram. For
these reasons, to the best of our knowledge, no observation of
Bose-Einstein condensation has been reported so far in $S=$ 1/2
tetramers.

In this work, we present experimental and theoretical evidence for
Bose-Einstein condensation in a crystal of \nit, a free radical-based
organic magnet \cite{barone_ab_1993} which behaves as weakly
interacting $S=$ 1/2 tetramers.  We show that the physics can be
described in terms of a fully rotational invariant system of quantum
spins without frustration.  When increasing the magnetic field, at low
temperature, we find the existence of three quantum phases.
We interpret two of them as having the physics of BECs.
In the low-field phase only the edge spins of each tetramer contribute
to the condensate, while in the high-fields phase, the order is
determined by the two central spins.  These unusual BECs are separated
by an incompressible state at half-magnetization that is a genuine
quantum phase, with half of the spins forming dimerized pairs, and the
other half aligned in the direction of the field.

The paper is organized as follows. The experimental details are given
in Section \ref{sec:exp_details} and the experimental results are
presented in Section \ref{sec:exp_results}. The latter includes, the
determination of the crystallographic structure
(\ref{subsec:cristal}), the characterization of the magnetic
properties from susceptibility and magnetization measurements
(\ref{subsec:susceptibility}), the evaluation of the magnetic
contribution to the specific heat (\ref{subsec:cv}), and the
determination of the temperature versus magnetic field phase diagram
(\ref{subsec:phase_diagram}). The theoretical evaluation of the
effective exchange interactions of the Heisenberg Hamiltonian is
presented in Section \ref{sec:exchange} and the determination of the
ground state of the system versus the applied magnetic field is given
in Section \ref{sec:dmrg}. Section \ref{sec:summary} concludes the
paper with a short summary.

\section{Experimental details}
\label{sec:exp_details}

The organic insulator
2-(2-Pyridyl)-4,4,5,5-tetramethyl-4,5-dihydro-1,H-imidazole-3-oxide-1-oxyl,
shortly called \nit, is part of the nitronyl nitroxide family.
Crystals of \nit\ were grown according to the method published in
References~\onlinecite{osiecki_studies_1968,ullman_stable_1970} and
single crystals up to $1\times1\times10$~mm have been obtained.  The
crystal structure was confirmed in a single-crystal X-ray diffraction
experiment performed on a Bruker Microstar X8/Proteum diffractometer
equipped with a Copper rotating anode delivering Cu K-alpha radiation
through multilayer Helios mirror optics. These data can be obtained
free of charge from the Cambridge Crystallographic Data Centre
(CCDC-1531994).

The magnetic susceptibility and magnetization were measured in a
commercial Quantum Design VSM SQUID magnetometer in the temperature
range from 1.8 to 300~K and magnetic fields up to 7~T, where the
sample was mounted with Apiezon N grease. For temperatures from 0.5 to
2~K, and magnetic fields up to 7~T we used a Quantum Design SQUID
magnetometer equipped with an iHelium3 option from IQUANTUM. Here the
sample was positioned in a Kapton tube and fixed with Teflon tape. We
also carried out measurements in pulsed magnetic fields up to 20~T in
a pumped $^4$He cryostat. Here, the sample was fixed inside a
compensated pick-up coil with Apiezon N grease.

The specific heat at ambient and under pressure was measured between
0.35 and 35~K in Quantum Design Physical Properties Measurement System
(PPMS) equipped with a $^3$He option and in magnetic fields up to
9~T. The specific heat under pressure was measured using a homemade
miniature CuBe pressure clamp \cite{miclea_investigation_2005} with a
small piece of lead as the pressure indicator. This pressure cell is
small enough to fit into the $^3$He insert of a PPMS. The
magnetocaloric measurements were carried out in a dilution
refrigerator equipped with a 20~T magnet. The sample was attached to a
sapphire chip with Apiezon N. This platform has a thermometer, and is
weakly coupled to a temperature regulated bath. The measurements were
then carried out by placing the sample at a specific point in the H-T
phase diagram and then a field ramp was started.  During the ramp, the
temperature of the bath was set to the sample temperature in a closed
loop, while temperature and field were recorded continuously.

\section{Experimental results}
\label{sec:exp_results}

\subsection{Crystallographic structure} 
\label{subsec:cristal}

\begin{figure}[htb!]
\begin{center}
  \includegraphics[width=0.48\textwidth]{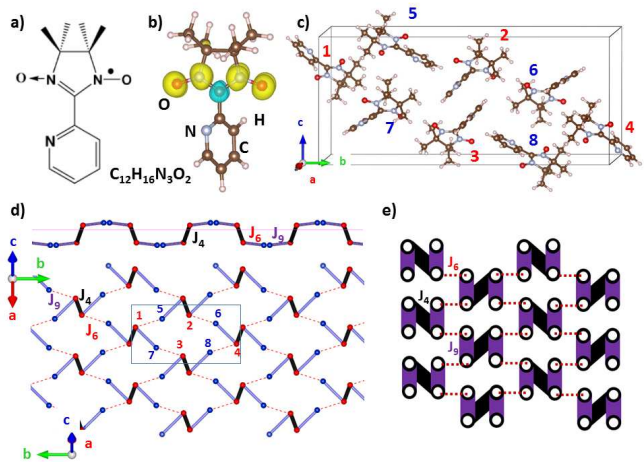}
 \caption{\label{fig:structure} (Color online)
   Atomic and magnetic structure of \nit.
   (a) Structural formula.
   (b) Atomic structure and isosurfaces of positive (yellow) and
   negative (blue) spin density ($\pm$ 0.003 e/\AA$\sp3$). Oxygen
   atoms are represented in red, carbon atoms in brown, nitrogen atoms
   in gray, and hydrogen atoms in light pink.
   (c) Monoclinic unit cell.
   (d) Lateral and top view of the 2D arrangement of the three leading
   magnetic interactions in the ${-1,0,2}$ plane: $J_4$ (black), $J_6$ (red), and $J_9$
   (violet).  The circles represent the C atom in the central
   O-N-C-N-O branch and the numbers correspond to those in (c). The
   solid lines outline a unit cell.
   (e) Topologically equivalent network of the magnetic lattice used
   for the DMRG calculations.}
\end{center}
\end{figure}

\nit\ crystallizes in the P2$_1$/c space group No. 14.  The chemical
and atomic structure of the isolated molecule is shown in
Figures~\ref{fig:structure}(a) and (b) and the monoclinic unit cell in
Fig.~\ref{fig:structure}(c).
It contains 264 atoms.  The lattice parameters are a = 6.1471 \AA, b =
30.0605 \AA, c = 12.9583 \AA, and $\beta = 100.269\sp{\circ}$.  There
are 8 molecules per unit cell \cite{barone_ab_1993} belonging to two
inequivalent groups of four molecules each (molecules numbered in red
1 to 4 and numbered in blue 5 to 8).

\subsection{Susceptibility and magnetization}
\label{subsec:susceptibility}

\begin{figure}[htb!]
\begin{center}
  \includegraphics[scale=0.5]{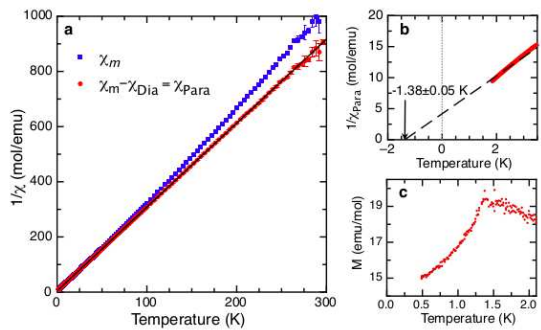}
 \caption{(Color online)
  (a) The magnetic susceptibility $\chi_{m}$ of a \nit\ polycrystal
   measured in an applied field of 1000~Oe is shown as blue squares.
   The paramagnetic susceptibility $\chi_\mathrm{Para} = \chi_{m} -
   \chi_\mathrm{Dia}$ obtained after the subtraction of
   $\chi_\mathrm{Dia}$ is shown as red circles. The solid line is a
   straight line fit of $\chi_\mathrm{Para}$ to a Curie-Weiss law.
  (b) Low-temperature region of $1/\chi_\mathrm{Para}$ vs. $T$. The
   dashed line is the Curie-Weiss fit.  (c) Magnetization of a single
   crystal for temperatures below 2~K measured in a applied field of
   50~Oe showing the behavior characteristic of antiferromagnetic
   order.}
\label{fig:Susceptibility}
\end{center}
\end{figure}

While the magnetism of metallic ions arises from unfilled atomic $d$
or $f$ orbitals, the magnetic moment in free radicals stems from
unfilled molecular orbitals. For each NIT-2Py molecule, there is one
unpaired electron that leads to a spin $S=1/2$ per molecule.
The inverse of the magnetic susceptibility $\chi$ measured on
polycrystalline \nit\ is shown in Fig.~\ref{fig:Susceptibility}(a).  A
deviation from the expected Curie-Weiss law is observed due to
significant diamagnetic contributions $\chi_{\mathrm{Dia}}$ (see
Fig.~\ref{fig:Susceptibility}(a)).  As \nit\ carries only one spin
$S=1/2$ per molecule, which contains a total of 33 atoms, the
diamagnetic contribution to the magnetic susceptibility from molecular
bonds is significant.  It has been subtracted requiring that the
remaining (paramagnetic) part would follow a perfect Curie-Weiss law
resulting in a $\chi_{\mathrm{Dia}}$ of $-131(2)~\mu$emu/mol. A value
which is of the same order of magnitude as the one that can be
calculated from tabulated values of Pascal's contributions from closed
molecular orbitals \cite{bain_diamagnetic_2008}.

Fitting a Curie-Weiss law to $\chi_{\mathrm{Para}} = \chi -
\chi_{\mathrm{Dia}}$ results in an effective moment of $1.71 \pm
0.01$~\mub\ per molecule, which is in agreement with the expected
value of $g\sqrt{(S(S+1)}$ for a $S=\frac{1}{2}$ spin from the
unpaired electron of each \nit\ molecule.  The Curie-Weiss temperature
of $\theta_\mathrm{CW}$ of $-1.38\pm0.05$~K (see
Fig.~\ref{fig:Susceptibility}(b)) points to antiferromagnetic
interactions between the \nit\ molecules.

Measurements on a single crystal of \nit\ along different
crystallographic directions showed no significant evidence for an
angular dependence of the magnetic susceptibility after we corrected
for sample geometry \cite{chen_demagnetizing_1991}. This is expected
for an organic compound such as \nit\ with small spin-orbit coupling.

Temperature dependent magnetization measurements at 50 Oe and below
2~K on a single crystal show a maximum around 1.4~K and a point of
inflection at 1.3~K (see Fig.~\ref{fig:Susceptibility}(c)) indicating
a possible antiferromagnetic transition at a characteristic
temperature similar to the Curie-Weiss temperature.

\begin{figure}[htb!]
\begin{center}
  \includegraphics[scale=0.3]{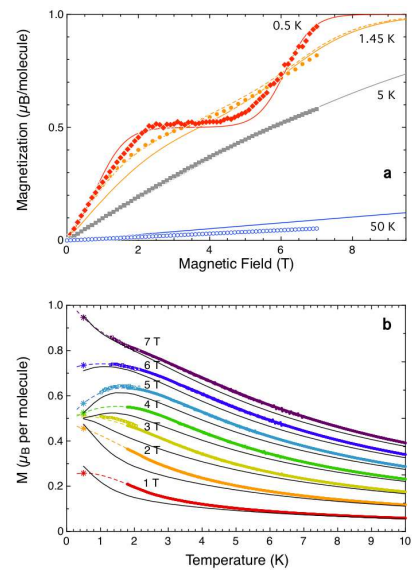}
 \caption{(Color online) (a) Magnetization of \nit\ as a function of
   the applied field. The data at 0.5~K is shown as solid diamonds,
   for 1.45~K as solid circles, for 5~K as filled squares, and for
   50~K as open circles.  The dashed line shows pulsed field
   magnetization at 1.43~K. The solid lines are obtained by minimizing
   the difference between the calculated magnetization obtained from
   diagonalizing Eq.~\ref{eq:Heisenberg} for a system of tetramers and
   the experimental results, as discussed in the last paragraph of
   Section~\ref{sec:exchange}.  (b) Magnetization of \nit\ as a
   function of temperature. The filled symbols are the data taken with
   a VSM SQUID, and the open symbols were taken with the $^3$He
   option. The solid lines are theoretical values calculated as
   described in (a).}
\label{fig:Magnetization}
\end{center}
\end{figure}

Magnetization isotherms of \nit\ at different temperatures are shown
in Fig.~\ref{fig:Magnetization}(a), where it can be seen that a
plateau at half of the saturation value begins to develop for
temperatures below 1.45~K and which is fully developed at 0.5~K.  The
observed saturation value corresponds to 1~\mub\ per molecule, as
expected for one free $S=\frac{1}{2}$ spin per \nit\ molecule.  The
dependence of the magnetization versus temperature at different
magnetic fields is shown in Fig.~\ref{fig:Magnetization}(b). The
convergence of the magnetization curves to 0.5 \mub\ at low
temperature for magnetic fields between 2 to 5~T corresponds to the
plateau at half the full magnetization.

While fractional plateaus are usually associated with quantum effects,
a simple possible explanation for the existence of this plateau could
be that one of the two crystallographically inequivalent groups of
molecules (see Fig.~\ref{fig:structure}(c)) form antiferromagnetic
dimers and the other ones behave as $S=\frac{1}{2}$ paramagnets
\cite{Rule2008,Aimo2009}.  This picture fails to be conclusive because
the magnetization increase between 0 and 2~T is slower than the
paramagnetic contribution 0.672~\mub$H/T$ indicating that other
antiferromagnetic interactions are also playing a significant role.

\subsection{Specific heat}
\label{subsec:cv}

\begin{figure}[htb!]
\begin{center}
  \includegraphics[scale=0.4]{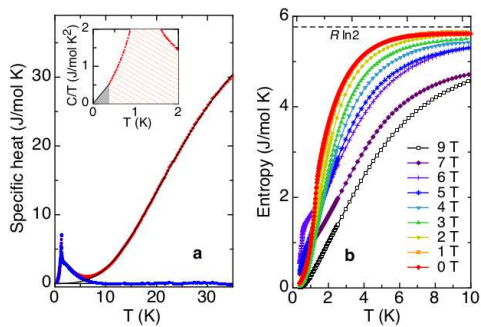}
 \caption{(Color online) (a) The measured specific heat $C_p$ of
   \nit\ in zero magnetic field, the phonon contribution from the
   lattice $C_{\mathrm{ph}}$, and the magnetic contribution $C_m = C_p
   - C_{\mathrm{ph}}$ are shown in black (solid line), red (diamonds)
   and blue (circles) respectively.  The inset shows the linear
   extrapolation of $C_p/T$ to zero temperature used to calculate the
   entropy. (b) Entropy associated with the phase transition in
   various magnetic fields for temperatures up to 10~K. The dashed
   line marks $S=R \ln{2}$, the value expected for a $S=\frac{1}{2}$
   per molecule.}
\label{fig:cp}
\end{center}
\end{figure}

The specific heat $C_p$ of \nit\ is shown in Fig.~\ref{fig:cp}(a)
for temperatures up to 35~K.  The magnetic contribution to the
specific heat is given by $C_m= C_p-C_{\mathrm{ph}}$, where
$C_{\mathrm{ph}}$ is the phonon contribution. This contribution was
estimated by fitting the specific heat above 12~K to a Debye model:
\begin{equation}
\label{eq:Debye}
            C_{\mathrm{ph}} = 9 N k_{\mathrm{B}} 
             \Bigl( \frac{T}{\theta_{\mathrm{D}}} \Bigr)^3
             \int_{0}^{\frac{\theta_{\mathrm{D}}}{T}} 
                       \frac{x^4 e^x}{(e^x - 1)^2} dx
\end{equation}
Here, $T$ is the temperature, $\theta_{\mathrm{D}}$ is the Debye
temperature, and $N$ the number of molecules. The fit resulted in a
$\theta_{\mathrm{D}}$ of 122~K. Such a low value of
$\theta_{\mathrm{D}}$ is expected in a system with weak bonds between
the molecules such as in a molecular crystal like \nit. The fit also
resulted in a $N$ of 2.1, which indicates that the two rings of the
molecule act as independent vibrational units. In zero applied fields,
$C_m$ features a sharp peak at $T_{c1}(0)=1.32$~K superimposed over a
large Schottky-like anomaly towards higher temperatures. This value of
$T_{c1}$ is the same temperature, at which we observe a point of
inflection in the magnetization, suggesting the presence of an
antiferromagnetic phase transition.

We calculated the magnetic entropy associated with the phase
transition by numerically integrating our specific heat data
$S=\int_0^T \frac{C_m}{T} dT$. The result of this integration is shown
in Fig.~\ref{fig:cp}(b). In order to be able to carry out this
integration, we extrapolated $\frac{C_m}{T}$ linearly to zero Kelvin,
as shown in the inset of Fig.~\ref{fig:cp}(b). The zero field
entropy shows that only a small fraction of the value of $S=R\ln 2$
expected for the magnetic entropy of a spin $S=\frac{1}{2}$ is
recovered just above the transition at 1.32 K. In order to fully
recover $S=R\ln 2$, we have to integrate up to 8~K, which indicates that
only a fraction of a spin $1/2$ is ordering in the transition.

\begin{figure}[htb!]
\begin{center}
  \includegraphics[scale=0.4]{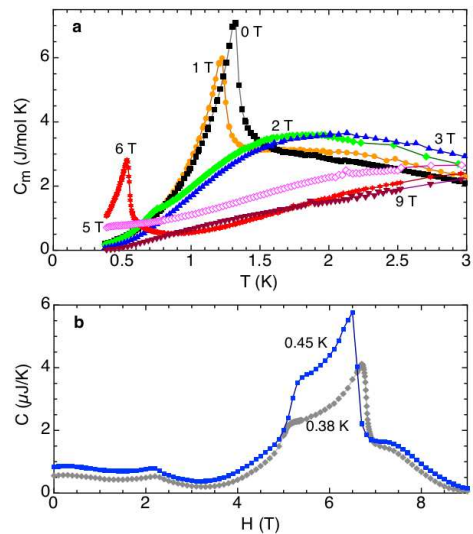}
 \caption{(Color online) (a) Magnetic contribution to the specific
   heat of \nit\ for various magnetic fields and temperatures from
   0.35 to 3~K. The zero field data shown as the solid squares shows a
   peak at 1.32~K. A field of 2~T almost completely suppresses this
   phase transition shown as the solid green diamonds, and by 3~T,
   shown as solid blue triangles, all that is left of the transition
   is very broad anomaly centered at 2~K. The red solid diamonds show
   the data at 6~T, where we see a second sharp peak associated with a
   second phase transition. This transition is fully suppressed by
   9~T. (b) Specific heat as a function of magnetic field measured at
   fixed temperatures. The data at 0.38~K, shown as solid diamonds,
   shows a first anomaly at about 2.2~T, followed by two more
   anomalies at 5.1 and 6.7~T.}
\label{fig:SpecificHeat}
\end{center}
\end{figure}

We have also carried out specific heat measurements in a number of
magnetic fields. The corresponding magnetic contributions are shown in
Fig.~\ref{fig:SpecificHeat}(a). The phase transition seen in zero
field is rapidly suppressed in a magnetic field. At a field of 2~T,
only a small peak is visible, whereas most of the weight of the
transition has merged with the Schottky-like anomaly centered at 2~K,
until the transition is completely suppressed at $H_{c1} \simeq
2.2$~T. 
Increasing the field further pushes the broad anomaly to higher
temperatures. At 6~T, a very sharp peak is observed, indicating the
presence of a second phase transition. This transition occurs only for
a limited field range, being absent at 5 and 7~T.  To map out this
second phase transition, we additionally carried out specific heat
measurements at fixed temperatures as a function of magnetic
field. Specific heat measurements versus magnetic field at 0.38~K
present anomalies at 2.2, 5.1 and 6.7~T confirming the existence of
the three phase transitions (Fig~\ref{fig:SpecificHeat}(b)). When
increasing the temperature the first anomaly shifts to lower magnetic
fields and the other two approach each other and finally disappear for
temperatures above the maximum critical temperature $T_{c2}(H=6$~T) of
0.53~K.

\subsection{Phase diagram}
\label{subsec:phase_diagram}

\begin{figure}[htb!]
\begin{center}
  \includegraphics[scale=0.5]{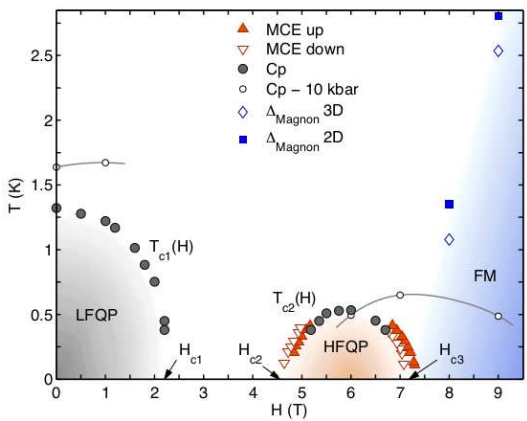}
 \caption{(Color online) Magnetic phase diagram $T$ vs $H$ deduced
   from sharp peaks in the specific heat $C_p$ (solid black circles),
   and magnetocaloric effect (MCE) measurements (red triangles, open
   for down swept fields, solid for fields swept up). Grey, red and
   blue regions represent the Low Field Quantum Phase (LFQP) , High
   Field Quantum Phase (HFQP) , and Ferromagnetic (FM) phases,
   respectively.  The solid squares (3D), and open diamonds (2D) are
   the size of the gap $\Delta$ extracted from a fit of the magnon
   excitation spectrum. The open circles are the phase boundaries of
   the BEC region of the phase diagram determined from specific heat
   with an applied pressure of 10~kbar.}
\label{fig:PhaseDiag}
\end{center}
\end{figure}

\begin{figure}[htb!]
\begin{center}
  \includegraphics[scale=0.5]{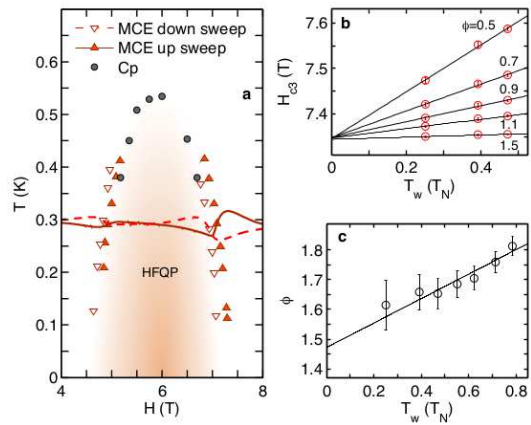}
 \caption{(Color online) (a) The solid black circles are the position
   of the anomaly from specific heat measurements.  The open red
   triangles indicate the position of the anomaly in temperature for
   fields being swept down, the solid red triangles are obtained from
   up sweeps.  The dashed (solid) red line presents a characteristic
   temperature-field trace for sweeping the magnetic field down (up).
   (b) Determination of the critical field $H_{c3}(0)$ by fitting the
   phase boundary $H_{c3}(T) -H_{c3}(0)\propto T^\phi$ for various
   values of the critical exponent and different temperature windows.
   (c) Determination of the critical exponent $\phi$ using the value
   of the critical field $H_{c3}$ of $7.345 \pm 0.003$~T found in
   (b).}
\label{fig:Magnetocaloric}
\end{center}
\end{figure}

The second anomaly seen in the specific heat forms a dome in the
$H-T$-phase diagram (see Fig.~\ref{fig:PhaseDiag}), which is
reminiscent of the Bose-Einstein condensation of magnons seen in
quantum paramagnets \cite{zapf_bose-einstein_2014}, and easy-plane
antiferromagnets with $U(1)$-rotational invariance around a
crystallographic axis \cite{Radu2005}.
In order to further explore the phase boundary of this field induced
dome, we carried out magnetocaloric measurements, which are shown in
Fig.~\ref{fig:Magnetocaloric}(a). Characteristic traces for fields
being swept up or down both show heating when the phase boundary is
crossed. This suggests that our sample and the thermal bath are in
equilibrium according to the discussion of magnetocaloric experiments
in Ref.~\onlinecite{zapf_bose-einstein_2014}.  We determined the phase
boundary as the midpoint between the two extrema of the
$H-T$-trace. The phase boundary determined from magnetocaloric
measurements is in fair agreement with the one determined from
specific heat measurements. The critical exponent $\phi$ of the upper
critical field $H_{c3}$ extracted from the results of the
magnetocaloric measurements  suggests that the field-induced order
is a Bose-Einstein condensation of magnons. The critical exponent is
related to the power law dependence of $H_{c3}(T) -H_{c3}(0) \propto
T^\phi$. Since the value of $\phi$ depends sensitively on $H_{c3}(0)$,
we followed the procedure laid out in Ref.~\onlinecite{Sebastian2005}
to obtain an accurate value for the critical exponent.

First, the critical field $H_{c3}(0)$ is determined by a fit to the
data for different temperature windows $T_w$ for various trial values
of $\phi$. The values of $H_{c3}(0)$ resulting from these fits are
shown in Fig.~\ref{fig:Magnetocaloric}(b). An accurate value of the
physical critical field is obtained by the extrapolation to an
infinitesimally small temperature window for each trial value of
$\phi$. Here, all the different extrapolations for different $\phi$
converge to $H_{c3}(0)=7.345 \pm 0.003$~T. Using this value, the
critical exponent $\phi$ was obtained through a similar extrapolation
to infinitesimally small temperature window, as shown in
Fig.~\ref{fig:Magnetocaloric}(c). The resulting value $\phi=1.47 \pm
0.09$ corresponds well to $\phi=1.5$ expected for a 3D Bose-Einstein
condensate of magnons
\cite{Affleck1991,Giammarchi_Tsvelik_2000,Nikuni2000a}.

\begin{figure}[htb!]
\begin{center}
  \includegraphics[scale=0.5]{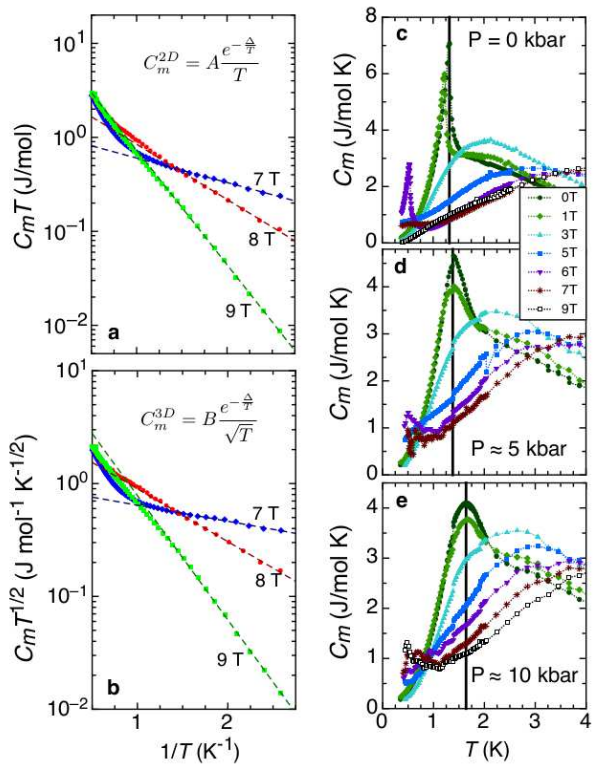}
 \caption{(Color online)(a) Semi-logarithmic plot of the magnetic
   specific heat $C_m$ of \nit\ presented as $C_m T$~vs.~$1/T$ for 7
   (diamonds), 8 (circles) and 9~T (squares). The dashed lines are
   fits to $C_m^{\mathrm{2D}}=A\frac{e^{-\frac{\Delta}{T}}}{T}$
   expected for 2D magnons. (b) Same data as shown in (a), but shown
   as $C_m \sqrt{T}$~vs.~$1/T$. The dashed lines as are fits to
   $C_m^{\mathrm{3D}}=B\frac{e^{-\frac{\Delta}{T}}}{\sqrt{T}}$. (c)
   Magnetic specific heat $C_m$ of \nit\ measured in a pressure cell
   at ambient pressure in the temperature range from 0.4 to 4~K. The
   solid line marks the position of the zero field anomaly. (d) $C_m$
   data for an applied pressure of 5~kbar. (e) $C_m$ data for 10~kbar
   of pressure.}
\label{fig:SpecificHeat2}
\end{center}
\end{figure}

The field dependence of $C_m$ at fixed temperatures (see
Fig.~\ref{fig:SpecificHeat}(b)) shows a Schottky-like anomaly for
fields above the upper critical field $H_{c3}$, indicating the
presence of a gap in the magnon spectrum of the field polarized
ferromagnetic phase. For the transition at $H_{c3}$ to be a
Bose-Einstein condensation this magnon gap needs to close at $H_{c3}$
 \cite{Radu2005}. To search for a magnon gap in \nit, we analyzed the
magnetic specific heat data for fields above $H_{c3}$. To extract the
size of the magnon gap $\Delta$ we tested contributions from a 2D, as
well as from a 3D magnon fluctuation spectrum. This analysis is shown
in Figs.~\ref{fig:SpecificHeat2}(a) and (b), respectively. Here, we
are following the example laid out for Cs$_2$CuCl$_4$, which in zero
field displays $XY$-antiferromagnetic order which is $U(1)$ invariant
around the $a$-axis \cite{Radu2005}.  Applying a magnetic field $H$
along the $a$-axis then breaks this $U(1)$ symmetry, as the transverse
spin component orders at $T_c$ \cite{Radu2005}. This leads to the
appearance of a Goldstone mode with a linear dispersion, which in the
case of Cs$_2$CuCl$_4$ in Ref.~\onlinecite{Radu2005} was interpreted
as the signature of a magnon Bose-Einstein condensation. We fitted our
7, 8 and 9~T data with
$C_m^{\mathrm{2D}}=A\frac{e^{-\frac{\Delta}{T}}}{T}$, which is
characteristic for a 2D magnon spectra, as well as
$C_m^{\mathrm{3D}}=B\frac{e^{-\frac{\Delta}{T}}}{\sqrt{T}}$, which is
characteristic for a 3D spectrum. Both curves fit our data equally
well in the available temperature and magnetic field range, and we are
unable to determine the dimensionality of the magnons in the field
induced ferromagnetic phase. The values of the gaps obtained from our
fits, which are very similar in size for both models, are shown in the
$H-T$-phase diagram shown in Fig.~\ref{fig:PhaseDiag}. For both
spectra, the resulting magnon gap $\Delta$ disappears at $H_{c3}$ at
zero temperature, as required for case of Bose-Einstein condensation.

The interactions in \nit\ are due to the overlap of the atomic
orbitals of the different molecules. As organic materials often show a
drastic change of their physical properties, see for example
Ref.~\onlinecite{Toyota2007}, such a the appearance of
superconductivity and charge- or spin-density wave transitions. We
measured specific heat of \nit\ with applied pressures of 0, 5 and
10~kbar in a number of applied fields. The results are shown in
Figs~\ref{fig:SpecificHeat2}(c), (d), and (e). While pressure somewhat
broadens the anomalies in the specific heat, the anomalies are still
visible. Pressure increase the temperature of the first dome, as well
as the second dome, but also pushes the maximum of the domes and the
upper critical field of the anomalies to higher fields, as shown in
Fig.~\ref{fig:PhaseDiag}.

The $H-T$-phase diagram of Fig.~\ref{fig:PhaseDiag} combines specific
heat and magnetocaloric data. For fields below $H_{c1}$ of $\approx
2.2$~T, we find a low field quantum phase from specific heat
measurements. For this range of magnetic fields, the magnetization
increases approximately linearly with applied field at 0.5~K. For
fields between $H_{c1}$, and $H_{c2}$, the magnetization shows a
plateau at half the saturation value at 0.5~K The occurrence of half
magnetization plateaus is rare, and we are only aware of two examples:
the spin 1 dimers Ba$_3$Mn$_2$O$_8$
\cite{Samulon2008,samulon_asymmetric_2009}, and the organic biradical
F$_2$PNNNO \cite{Tsujii2005,Hosokoshi1999,Bostrem2010}. For fields
above $H_{c2}$ the magnetization increases again and saturates at
$H_{c3}$, where the specific heat and magnetocaloric results indicate
a phase boundary. This suggests, that \nit\ displays two field induced
Bose-Einstein condensations. This is also borne out by the vanishing
magnon gap $\Delta$ at $H_{c3}$, which was extracted from specific
heat data in magnetic fields larger than $H_{c3}$.

\section{Calculation of the effective exchange interactions}
\label{sec:exchange}

To understand the magnetic order at the origin of the phase
transitions one needs to determine the leading magnetic interactions
between the \nit\ molecules.
Due to the negligible anisotropy we have assumed that the magnetic
properties can be described by a rotational invariant Heisenberg
Hamiltonian:
\begin{equation}
\label{eq:Heisenberg}
\hat{H} = \hat{H}_0 + \sum_{i > j} J_{ij} \; \mathbf{\hat{S}_{i}} \cdot
\mathbf{\hat{S}_{j}},
\end{equation}
where $\hat{H}_0$ is the spin-independent part of the Hamiltonian,
$J_{ij}$ are the magnetic couplings, and $\mathbf{\hat{S}_{i}}$ and
$\mathbf{\hat{S}_{j}}$ are the $S=\frac{1}{2}$ spin operators
localized on the \nit\ molecules at site $i$ and $j$, respectively.
Unfortunately, it is difficult to see an obvious arrangement of the
molecules which can be used to predict the relative strength of the
exchange interactions by inspection of the crystal structure.

Moreover, compared to transition metals oxide based quantum magnets
\cite{saul_magnetic_2011,radtke_interplay_2010,saul_density_2014}, the
spin polarization in \nit\ is highly delocalized on the O-N-C-N-O
branch in the center of the molecule (see Fig.~\ref{fig:structure}(b))
like in other members of the family
\cite{zheludev_neutron_1995,hirel_cyano_2002}. Hence, the interactions
are expected to have a rather extended range.
For this reason, we have calculated 13 different interactions up to 
intermolecular distances of 9.404~\AA\ (see Table
\ref{table:exchange}).
To classify the exchange interactions we used the distance between the
central C atoms in the O-N-C-N-O branch of each molecule.

The calculations were performed using a broken-symmetry formalism,
i.e., by mapping total energies corresponding to various collinear
spin arrangements within a supercell onto the Heisenberg Hamiltonian
of Eq.~\ref{eq:Heisenberg}.
For the calculations we have used the \textsc{Quantum Espresso}
\cite{Giannozzi2009} code based on density functional theory,
ultrasoft pseudopotentials and the PBE
functional~\cite{perdew_generalized_1996} with a plane-wave and
charge-density cutoff of 80 Ry and 320 Ry, respectively.  We have used
a 4x1x2 Monkhorst-Pack~\cite{Monkhorst1976} grid for the first
Brillouin zone sampling of the 264 atoms monoclinic
1$\times$1$\times$1 unit cell and adapted equivalent samplings for the
double 2$\times$1$\times$1 and 1$\times$1$\times$2 or the quadruple
2$\times$1$\times$2 super-cell calculations. A full relaxation of the
internal coordinates of the 264 atoms has been performed in the
1$\times$1$\times$1 cell. The same relaxed coordinates have been
consistently used to construct the supercells.  The different
supercells were needed to distinguish the exchange interactions
between a molecule and two different molecules which are connected by
the translation symmetry if the 1$\times$1$\times$1 unit cell is used.
Only the 1056 atoms 2$\times$1$\times$2 unit cell allows to calculate
separately the 13 interactions. The 264 atoms 1$\times$1$\times$1 unit
cell, whose total energy can be written as :
\begin{eqnarray}
\nonumber
E\sp{111} & = & E_0 +  \frac{1}{4}\ 
        [ 4\ (J_1 + J_{1}^\prime) + b_2\ (J_2 + J_4 + J_7 + J_{11}) \\
\nonumber
         & + & b_3\ (J_3 + J_{10}) + b_5\ (J_5 + J_9) \\
         & + & b_6\ (J_6 + J_{12}) + b_8\ J_8 ]      
\label{eq:ene111}
\end{eqnarray}
neither allow to calculate $J_1$ and $J_{1}^\prime$ nor to separate
$J_2$, $J_4$, $J_7$, and $J_{11}$; $J_3$ and $J_{10}$; $J_5$ and
$J_9$; and $J_6$ and $J_{12}$.
Similar limitations arise with the 528 atoms 1$\times$1$\times$2 and
2$\times$1$\times$1 unit cells, whose total energies are :
\begin{eqnarray}
\nonumber
E\sp{112} & = & 2\ E_0 +  \frac{1}{4}\ 
           [ 8\ (J_1 + J_{1}^\prime) + c_2\ (J_2 + J_7 ) \\
\nonumber
         & + & c_4\ (J_4 + J_{11}) + c_3\ J_3 + c_5\ J_5 \\
         & + & c_6\ (J_6 + J_{12}) + c_8\ J_8 + c _9\ J_9 + c_{10}\ J_{10} ] 
\label{eq:ene112}
\end{eqnarray}
and
\begin{eqnarray}
\nonumber
E\sp{211} & = & 2\ E_0 +  \frac{1}{4}\ 
         [ d_1\ J_1 + d_1^\prime\ J_{1}^\prime + d_2\ (J_2 + J_4 ) \\
\nonumber
         & + & d_3\ J_3 + d_5\ J_5 + d_6\ J_6 + d_7\ (J_7 + J_{11}) \\
         & + & d_8\ J_8 + d _9\ J_9 + d_{10}\ J_{10} + d_{12}\ J_{12} ]
\label{eq:ene211}
\end{eqnarray}
The coefficients $b_j$, $c_j$, and $d_j$ depend on the spin
arrangements of the molecules.

Two different calculation procedures have been used to calculate the
effective exchange interactions.  The first procedure uses a
least-squares minimization of the difference between the DFT and Ising
relative energies to obtain a numerical evaluation of the couplings.
The second procedure allows to calculate separately the effective
exchange interaction. For example, the interaction between spin $i$
and $j$ can be evaluated from:
\begin{equation}
   J_{ij} = E_{ij}(\uparrow \uparrow)   + E_{ij}(\downarrow \downarrow)
          - E_{ij}(\uparrow \downarrow) - E_{ij}(\downarrow \uparrow),
\end{equation}
where $E_{ij}(\sigma_i,\sigma_j)$ are the four spin configurations
where the spins $i$ and $j$ take the values up or down while all the
other spins are kept up\cite{Xiang2013}. $J_{ij}$ could be a single or
a sum of exchange interactions depending on the size of the unit cell
used to calculate the total energies.

As the 1$\times$1$\times$1 unit cell contains 8 molecules there is a
total of 256 distinct spin configurations. However, taking crystal and
spin reversal symmetries into account this number can be reduced to
39. The application of the least-squares minimization procedure to
this unit cell gives a first estimation of the exchange interactions
(in units of K):
\begin{center}
\begin{tabular}{lcS[table-format=3.1]}
$J_2 + J_4 + J_7 + J_{11}$ & = & 12.5 \\
$J_3 + J_{10}$             & = & -0.5 \\
$J_5 + J_9$                & =  & 7.5 \\
$J_6 + J_{12}$             & =  & 8.6 \\
$J_8$                      & = & -0.2 \\
\end{tabular}
\end{center}
The second procedure\cite{Xiang2013} was used with the larger unit
cells to evaluate $J_1$ and $J_1^\prime$ and separate the exchange
interactions.  All the calculations gave consistent values of the
exchange interactions with an overall error of $\pm\ 0.1$\ K. For
example, with the 2$\times$1$\times$1 unit cell we obtain :
\begin{center}
\begin{tabular}{lcS[table-format=3.1]}
$J_2 + J_4    $ & = &  11.3 \\
$J_7 + J_{11} $ & = &   1.2  \\
$J_5          $ & = &  -0.5  \\
$J_9          $ & = &   8.1  \\
$J_6          $ & = &   6.2  \\
$J_{12}       $ & = &   2.3  \\
\end{tabular}
\end{center}
whose corresponding sums are in good agreement with the values
obtained from the single unit cell.  Similarly, with the
1$\times$1$\times$2 unit cell we get :
\begin{center}
\begin{tabular}{lcS[table-format=3.1]}
$J_2 + J_7     $ & = &  0.5  \\
$J_4 + J_{11}  $ & = & 12.0  \\
\end{tabular}
\end{center}
in agreement with the above estimations. A summary of the calculated
exchange interactions is shown in Table \ref{table:exchange}.
\begin{table}[htb!]
\begin{center}
\begin{ruledtabular}
\begin{tabular}{cccc}
  & d$_\text{C-C}$ [\AA] & Equivalent group  & $J_\textit{i}$ [K] \\
\hline
$J_1$          & 6.15 & 1-1 & $-$0.9 \\
$J_1^\prime$   & 6.15 & 2-2 & $-$2.9 \\
$J_2$          & 6.43 & 1-1 & $-$0.6 \\
$J_3$          & 6.68 & 2-2 & $-$0.5 \\
$J_4$          & 7.00 & 1-1 & $+$11.9 \\
$J_5$          & 7.40 & 1-2 & $-$0.6 \\
$J_6$          & 7.86 & 1-2 & $+$6.2 \\
$J_7$          & 7.94 & 1-1 & $+$1.1 \\
$J_8$          & 8.08 & 1-2 & $-$0.2 \\
$J_9$          & 8.20 & 1-2 & $+$8.1 \\
$J_{10}$       & 8.26 & 2-2 & $+$0.0 \\
$J_{11}$       & 8.63 & 1-1 & $+$0.1 \\
$J_{12}$       & 9.40 & 1-2 & $+$2.4 \\
\end{tabular}
\end{ruledtabular}
\caption{Effective exchange interactions.  The thirteen interactions
  calculated in this work between the \nit\ molecules obtained using
  density functional theory are listed in the first column.
  The distances in the second column are measured between the
  C atoms in the O-N-C-N-O branch of each molecule.
  The third column gives the equivalent groups of the molecules
  associated with the corresponding exchange interaction.
  In the last column, the effective interactions are given in units of
  K. A positive value is associated to an antiferromagnetic
  interaction.}
\label{table:exchange}
\end{center}
\end{table}

It is interesting to note that in spite of the fact that most of the
interactions have non-zero values, the three leading couplings are all
antiferromagnetic with positive values, namely, $J_4 = 11.9$, $J_6 =
6.2$, and $J_9 = 8.1$~K.
The strongest interaction $J_4$ is represented by the thick black
lines in Fig.~\ref{fig:structure}(d) and connects molecules, which are
related by symmetry (1 and 4 or 2 and 3 as labeled in
Fig.~\ref{fig:structure}(c)).
If one sets $J_6$ and $J_9$ to zero the equivalent magnetic lattice
would correspond to dimers on one of the two sublattices formed by one
of the groups of four crystallographically equivalent molecules and
isolated paramagnets on the other.
If one keeps the second-largest interaction, $J_9$, represented by
violet lines in Fig.~\ref{fig:structure}(d), the system becomes an
ensemble of independent tetramers consisting of four $S=\frac{1}{2}$
moments.
When the third-strongest term $J_6$ is added, the magnetic structure
forms a corrugated 2D lattice of interacting tetramers, as shown in
Fig.~\ref{fig:structure}(d).  This family of planes is indexed by
\{-1,0,2\}.
Despite the complexity of the structure and the large number of
couplings the system is not frustrated, and it is possible to satisfy
the conditions for an antiferromagnetic $S_\text{tot} = 0$ N\'eel
order (see Fig.~\ref{fig:neel}), in agreement with experiments.

Magnetization isotherms calculated by exact diagonalization for a
system of four coupled tetramers show a good qualitative agreement
with the experimental data in spite of a systematic shift of the
critical fields and temperatures to larger values.  The reason for
this difference lies in the known overestimation of the exchange
interactions when a semilocal functional is used
\cite{Mazurenko2014,saul_density_2014}.  A quantitative match with the
experiments requires smaller values of the exchange interactions.

 The order of magnitude of the intra-tetramer exchange interactions
 ($J_4$ and $J_9$) can be obtained by comparing the two critical
 fields
$E_\text{ST} (J_4, J_9)$ and
$E_\text{TQ} (J_4, J_9)$ (equations (\ref{eq:S2T}) and
  (\ref{eq:T2Q})), corresponding to the stabilization of the triplet
  and quintuplet ground states of the isolated tetramer, to the values
  of the magnetic field at the center of the domes in the $T$ vs $H$
  phase diagram shown in Fig.~\ref{fig:PhaseDiag}.
The intensity of the inter-tetramer interaction ($J_6$) can be
estimated from the width of the domes at zero temperature.  The
critical fields of about 1.1 T and 5.8 T and a half width 1.25 T (see
Fig.~\ref{fig:PhaseDiag}) give a rough estimate of $J_4 = 5.7$, $J_6 =
1.7$, and $J_9 = 3.3$~K.
A more precise estimation can be obtained by a least-squares
minimization of the differences between the experimental and
theoretical magnetization obtained by exact diagonalization of the
Heisenberg Hamiltonian given in Eq.~\ref{eq:Heisenberg}.  With this
procedure we obtained $J_4 = 6$, $J_6 = 1$, and $J_9 = 2.8$~K. These
values has been used in the rest of the work and for the solid lines
in Fig.~\ref{fig:Magnetization}.

\section{Calculation of the ground state versus magnetic field}
\label{sec:dmrg}

In order to determine the ground-state of this system in an applied
magnetic field we performed density matrix renormalization group
calculations (DMRG) \cite{White1992,White1993}.
For clarity and convenience, we have placed the tetramers on the
vertices of a square lattice, as shown in Fig.~\ref{fig:structure}(e).
The calculations were performed on cylinders of different aspect
ratios.
The antiferromagnetic exchange between tetramers $J_6$ is smaller by
at least a factor of three compared to the ones within tetramers,
$J_4$ and J$_9$.  In zeroth-order approximation we can consider the
ground-state to be a crystal of singlets.  Nevertheless, these
interactions are very important, since they are responsible for
establishing long-range magnetic order.  The weakly entangled nature
of our model makes it amenable to DMRG calculations, which have
already proven very successful in unveiling the magnetic phases of the
Shastry-Sutherland compound SrCu$_2$(BO$_3$)$_2$
\cite{Jaime2012,Haravifard2016,Matsuda2013}.  Simulations at zero
field yield a small but finite singlet-triplet gap of
$0.38$~K. Moreover, the ground-state energy per tetramer $E_0 =
-0.9787\ J_4$ is very close to the value for an isolated tetramer
$-0.9675\ J_4$, indicating that the ground-state is a crystal of
tetramers without long-range antiferromagnetism (the dependence of the
ground state energy is shown in Fig.~\ref{fig:dmrg_vs_d}).  However,
it is possible that inter-layer or additional interactions could close
the gap and establish true long-range order. We notice that the zero
field critical temperature is $T_{c1}(0) =1.32$~K, so it is possible
that the material is very close to a quantum critical regime
separating a magnetically ordered state from a crystal of tetramers.
At the magnetization plateau at $m=1/2$ the Heisenberg contribution to
the ground-state energy is $E_{1/2} =-0.8038\ J_4$, whereas the one of
islated tetramers is $-0.8017\ J_4$.  Therefore, the plateau can also
be described as an incompressible crystal of tetramers, in which the
spins sitting at the edges on the weak bonds are fully polarized in
the direction of the field, and the two central spins form a tightly
bound dimer.  Explicitly, the wave function of a single tetramer at
half magnetization can be written as:

\begin{equation}
  |\mathrm{g.s.}\rangle_{m=1/2} = \alpha|\psi_1\rangle +
  \beta|\psi_2\rangle, 
\end{equation}
with 
\begin{equation}
  |\psi_1\rangle=1/\sqrt{2}\left[|\uparrow \uparrow \uparrow
  \downarrow\rangle-|\downarrow \uparrow \uparrow
  \uparrow\rangle\right],
\end{equation}
and
\begin{equation}
|\psi_2\rangle=1/\sqrt{2}\left[|\uparrow \uparrow \downarrow
  \uparrow\rangle-|\uparrow \downarrow \uparrow
  \uparrow\rangle\right],
\end{equation}
describing a singlet between the two edge spins and between the
central spins, respectively. In our case we find $\beta^2=0.95$,
meaning that the latter carries almost all the weight. Although we
assume this picture of decoupled tetramers to simplify the description
of the problem, in reality the DMRG simulations indicate that the
moment of the edge spins is $\langle S^z \rangle = 0.48$ and finite
but very small correlations $\langle S^+_iS^-_j\rangle \sim 10^{-3}$
connect nearby tetramers.

The BEC regime is realized both between zero and the lower critical
field $H_{c1}$ and between the end of the plateau at $H_{c2}$ and full
polarization at $H_{c3}$, corresponding to the grey LFQP and red
HFQP regions in Fig.~\ref{fig:PhaseDiag}, respectively.  As the
magnetic field increases, the edge spins start canting in the
direction of the field, simultaneously establishing a correlated state
with long-range order in the transverse plane. In bosonic language,
the edge spins form a superfluid with off-diagonal long-range order,
while the central spins remain dimerized. A similar behavior is found
above the plateau, with the central spins canting in the direction of
the field, while the edge spins remain fully polarized.

In order to characterize the different field-induced phases we
calculated the longitudinal and transverse spin-structure factors,
defined as:
\begin{eqnarray}
    S^{z}({\mathbf{q}}) &=& \frac{1}{N}\sum_{ij} \langle S^z_i S^z_j 
               \rangle e^{i\mathbf{q} \dot (\mathbf{r}_i-\mathbf{r}_j)}, \\
    S^{+-}({\mathbf{q}}) &=& \frac{1}{N}\sum_{ij} \langle S^+_i S^-_j 
               \rangle e^{i\mathbf{q} \dot
               (\mathbf{r}_i-\mathbf{r}_j)}, 
\end{eqnarray}
where the $z$ direction is chosen along the applied magnetic field and
the spin coordinates $\mathbf{r}_i$ and momenta $\mathbf{q}$ are the
ones of the topologically equivalent square lattice mentioned above.
In the $S^z$ basis, these quantities measure diagonal and off-diagonal
long-range order, respectively.

\begin{figure}[htb!]
\begin{center}
    \includegraphics[width=0.45\textwidth]{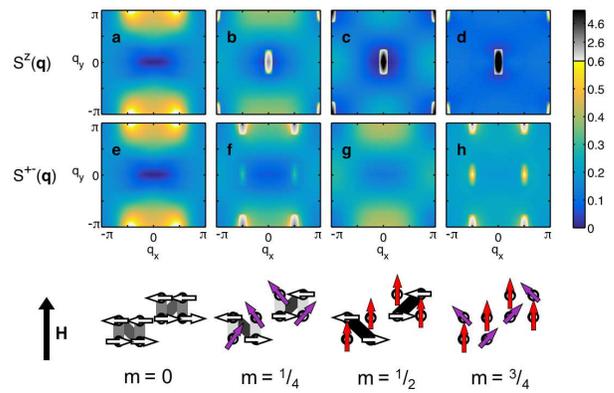}
  \caption{ (Color online) Structure factor and magnetic order in
    \nit\ calculated by DMRG in the Brillouin zone of the square
    lattice for a cylinder of size $L_x \times L_y = 32 \times 8$ (see
    text).
    (a)-(d) Longitudinal and
    (e)-(h) transverse components of the spin structure factor for
    different values of magnetization $m =
    0,\ 1/4,\ 1/2,\ \text{and}\ 3/4$. The corresponding magnetic order
    is sketched below each column. Red spins align along the field
    direction, while violet spins have a component along the field
    direction and order in the plane transverse to the field.
    White/empty arrows represent spins in a rotational invariant
    quantum superposition for tetramers.  The gray scale in the
    bonds indicates the relative strength of the correlations, with
    black representing a strong dimer.}
  \label{fig:sq}
\end{center}
\end{figure}

\begin{figure}[htb!]
\begin{center}
  \includegraphics[width=0.4\textwidth]{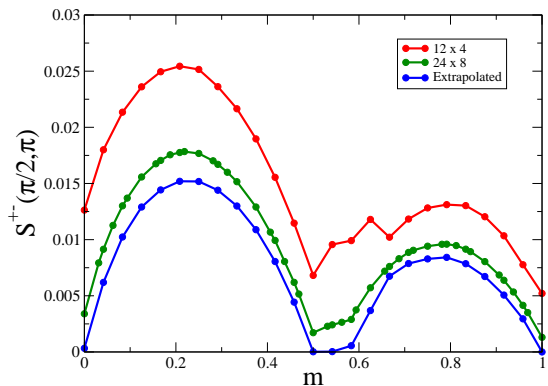}
  \caption{ (Color online) Transverse structure factor
    $S^{+-}(\pi/2,\pi)$ for two system sizes with the same aspect
    ratio, $N=L_x \times L_y = 12 \times 4$ and $24 \times 8$. Results
    in the thermodynamics limit were obtained through a linear
    extrapolation in $1/N$. Numerical errors are smaller than the
    symbol size. }
  \label{fig:spi2pi}
\end{center}
\end{figure}

Results for different magnetization values are shown in
Fig.~\ref{fig:sq}.  Panels (a)-(d) display the longitudinal component
$S^{z}$, while panels (e)-(h) show the transverse  $S^{+-}$ component.
Note that the unit cell used for these calculations is a single spin
on a square lattice (Fig.~\ref{fig:structure}(e)).
The corresponding order is sketched below. At $m=0$ the correlations
do not display a sharp peak, and they are almost evenly distributed
along the $\mathrm{q_y}=\pm \pi$ axes. This result can be easily
recovered by considering a crystal of tetramers in their singlet
ground state.
At $m=1/2$ we similarly can reproduce the measured quantities by
assuming a crystal of triplets.  The edge spins are fully polarized,
as reflected in the peaks of the longitudinal structure factor at
$\mathrm{q}=(\pi,\pi)$ (see Fig.~\ref{fig:sq}).  The peak at
$\mathrm{q}=(0,0)$ is proportional to the total magnetization
squared. In the transverse direction, we do not observe a sharp peak,
and our results describe a valence-bond solid, or crystal of dimers.
At $m=1/4$ and $m=3/4$ the off-diagonal correlations show sharp
additional peaks at $\mathrm{q}=(\pi/2,\pi)$, indicating the onset of
long-range order (see Fig.~\ref{fig:sq}).  It is important to
highlight that this regime cannot be explained in terms of isolated
tetramers, and emerges as an effect of correlations and due to the
inter-molecule interactions. In this sense, neither dimers nor edge
spins are fully disentangled.
In order to determine the existence of off-diagonal order in the
thermodynamic limit we perform a finite size scaling of the structure
factor, shown in Fig. \ref{fig:spi2pi}, We carried out a linear
extrapolation in $1/N$ using cylinders with the same aspect
ratio. Results indicate a finite window around $m=1/2$ where the
off-diagonal correlations vanish, suggesting the existence of a new
phase with a coexistence of fully polarized spins and a disordered
state resulting from the ``melting'' of the valence bond solid.

\section{Summary}
\label{sec:summary}

Our experimental data backed by comprehensive theoretical and
numerical analysis demonstrate a rich and unconventional magnetic
behavior in the organic molecular crystal \nit\ with field-induced
phases that can only be interpreted in term of a quantum-mechanical
description.
Specific heat and magnetocaloric measurements indicated the presence
of two domes in $H-T$-phase diagram: At zero field \nit\ shows an
antiferromagnetic phase transition. However, the entropy associated
with this phase transition is only a fraction of $R \ln2$, indicating,
that the ground state of this crystal of spin 1/2 carrying molecules
is quantum mechanical in nature.  An applied field suppresses this
phase transition at a critical field $H_{c1}$ of 2.2~T.  This is the
same field at which the magnetization measured at 0.5~K becomes field
independent and shows a plateau at half the saturation value up to a
$H_{c2}$ of 4.5~T, where a second anomaly appears in the specific
heat. Here the magnetization starts to increase again linearly up to
saturation value of 1~$\mu_{\mathrm{B}}$ to saturate at a field of
$H_{c3}$ of 7.3~T, where the second anomaly in specific heat
disappears. The exponent $\phi$ of the power-law behavior $H_{c3}(T)
-H_{c3}(0) \propto T^\phi$ at $H_{c3}$ of this second dome in the
$H-T$-phase diagram corresponds to the value expected for a
Bose-Einstein condensation of magnons. This is supported by the magnon
gap $\Delta$ we see in the specific heat for magnetic fields above
$H_{c3}$, which closes at $H_{c3}$.

In order to be able to propose an effective model of the interactions
in \nit, we carried out a series of total energy calculations in the
so-called broken symmetry formalism, where the spins on the molecules
are polarized by hand. Due to lack of spin-orbit interaction in \nit,
the total energies can be mapped directly to the rotational invariant
Heisenberg Hamiltonian of Eq.~\ref{eq:Heisenberg}. By using super
cells of up to $2\times1\times2$ we were able to identify the
different exchange interactions between neighboring molecules. As
listed in Table~\ref{table:exchange} we found that the leading
interactions are all antiferromagnetic. The minimal magnetic model
obtained by mapping the coordination and strength of the interactions
back to the structure consists in spin 1/2 tetramers, which form a
corrugated 2D lattice parallel to the \{-1,0,2\} set of
crystallographic planes, as shown in Fig.~\ref{fig:structure}(d). The
strength of the interactions obtained from the broken symmetry
formalism are comparable to the values which result from fitting the
exchange constants to the magnetization data of \nit, as shown
Fig.~\ref{fig:Magnetization}(a).

Having established the minimal magnetic model, we carried out DMRG
calculations on finite but large systems and determined the magnetic
phase diagram
To summarize the qualitative picture that emerges from our results and
analysis, we find a low field BEC formed by the spins at the ends of
the tetramers, with the two spins in the middle strongly entangled
into dimers. The high field BEC is formed by the central spins, with
the ones at the edges practically fully polarized.  The high field BEC
is qualitatively similar to TlCuCl$_3$ \cite{Tanaka2001,Ruegg2003},
since right above the plateau the system basically consists of a
crystal of dimers, and can be described in the same language with the
(practically polarized) edge spins mediating the interactions between
the singlets.  Unlike most quantum magnets that realize a classical
``up-up-up-down'' order in the half-magnetization plateau,
\nit\ exhibits a true quantum state, similar to the one reported in
CdCu$_2$(BO$_3$)$_2$ \cite{Hase2009, Janson2012}, formed by a
valence-bond solid coexisting with fully polarized spins.  We hope
that \nit\ can become a new exciting playground to realize novel
states and study quantum phase transitions, for instance under
chemical doping or hydrostatic pressure.

\begin{acknowledgments}
The research at UdeM received support from the Natural Sciences and
Engineering Research Council of Canada (Canada), Fonds Qu\'eb\'ecois
de la Recherche sur la Nature et les Technologies (Qu\'ebec), and the
Canada Research Chair Foundation.  Part of this work was supported by
HLD at HZDR, a member of the European Magnetic Field Laboratory
(EMFL). N.G. thanks for the hospitality of the Max Planck Institute
for Chemical Physics of Solids. A.F. acknowledges the U.S. Department
of Energy, Office of Basic Energy Sciences, for support under grant
DE-SC0014407.  A. S.  and A.F. thank M. Jaime, C. Batista and G.
Radtke for fruitful discussions. D.L. and C.R. are grateful for
funding by the Centre National de la Recherche Scientifique (CNRS) for
collaborative research (PICS 2015-2017)
\end{acknowledgments}

\appendix

\section{Isolated tetramer}

The Hamiltonian of an isolated tetramer with interactions $J_4$ and
$J_9$ is
\begin{equation}
   H_{tetra} = J_9 (S_1 \cdot S_2 + S_3 \cdot S_4 ) + J_4 (S_2 \cdot S_3 )
\end{equation}
The system has two singlets ($S=$ 0), three triplets ($S=$ 1) and one
quintuplet ($S=$ 2) eigenstates whose energies are \cite{Hase1997} :
\begin{equation}
E_1\sp{S} = -\frac{J_9}{2}-\frac{J_4}{4} 
            - \frac{\sqrt{J_4^2 - 2 J_4 J_9 + 4 J_9^2}}{2}
\end{equation}
\begin{equation}
E_2\sp{S} = -\frac{J_9}{2}-\frac{J_4}{4} 
            + \frac{\sqrt{J_4^2 - 2 J_4 J_9 + 4 J_9^2}}{2}
\end{equation}
\begin{equation}
E_1\sp{T} = -\frac{J_4}{4} - \frac{\sqrt{J_4^2 +  J_9^2}}{2} 
\end{equation}
\begin{equation}
E_2\sp{T} = -\frac{J_4}{4} + \frac{\sqrt{J_4^2 +  J_9^2}}{2} 
\end{equation}
\begin{equation}
E_3\sp{T} = -\frac{J_9}{2} + \frac{J_4}{4}
\end{equation}
\begin{equation}
E_1\sp{Q} = \frac{J_9}{2} + \frac{J_4}{4} 
\end{equation}
With antiferromagnetic (positive) interactions, at zero magnetic
field, the ground state is the $E_1\sp{\text{S}}$ singlet.
With and applied magnetic field at zero temperature a first jump in
the magnetization arises when the $S_z = 1$ component of the lowest
energy triplet becomes the ground state at :
\begin{equation}
   H_{\text{ST}} = \frac{1}{g\ \mu_\mathrm{b}} (E_1\sp{\text{T}} - E_1\sp{\text{S}})
   \label{eq:S2T}
\end{equation}
and a second jump when the $S_z = 2$ component of the quintuplet
crosses the $S_z = 1$ energy of the triplet :
\begin{equation}
   H_{\text{TQ}} = \frac{1}{g\ \mu_\mathrm{b}} (E_1\sp{\text{Q}} - E_1\sp{\text{T}})
  \label{eq:T2Q}
\end{equation}

\section{Density matrix renormalization group calculations}

\begin{figure}[htb!]
  \includegraphics[width=0.3\textwidth]{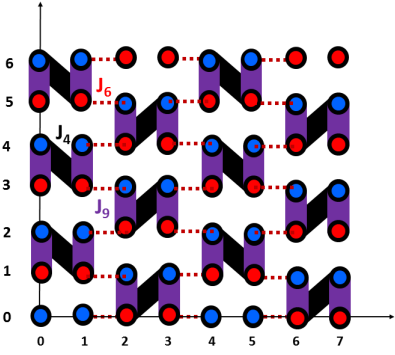}
\caption{Antiferromagnetic N\'eel order compatible with the calculated
  exchange interactions.}
\label{fig:neel}
\end{figure}

As described in the text, the geometry of the problem was mapped onto
a system of spins at the vertices of a square lattice.
Antiferromagnetic N\'eel order is compatible with the calculated
exchange interactions as can be seen in Fig.~\ref{fig:neel} where the
magnetic order is represented in the topologically equivalent lattice
and magnetic unit cell used for DMRG calculations.
The blue and red circles represent opposite projections of the
magnetic moment along $z$.  There is no frustration when the three
leading magnetic interactions between the \nit\ molecules $J_4$
(black), $J_6$ (red), and $J_9$ (magenta) are all
antiferromagnetic. This magnetic order corresponds to a 4 $\times$ 2
superstructure with respect to the underlying square lattice. It would
appear as peak at $\mathrm{q}=(\pi/2,\pi)$ in the spin structure
factor.

\begin{figure}[htb!]
   \includegraphics[width=0.35\textwidth]{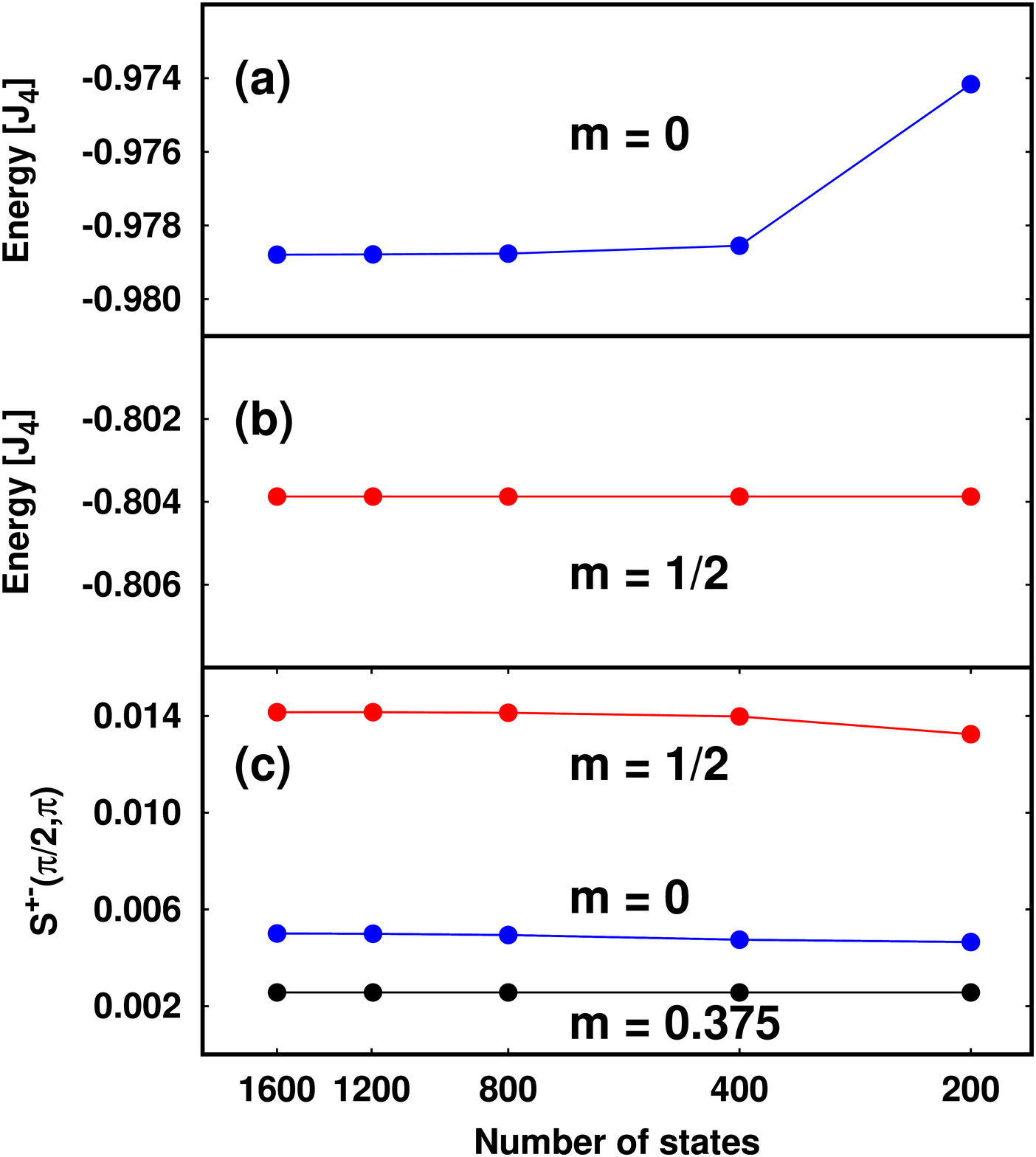}
   \caption{Dependence of the DMRG calculations with the number of
     states $d$ for a cylinder of size $L_x \times L_y = 16 \times 8$.
     Ground state energy per tetramer $E(d)$ (in units of $J_4$) and
     magnetization equal to (a) $m=0$ and (b) $m=1/2$.  Off-diagonal
     structure factor (c) $S^{+-}(\pi/2,\pi)$ for $m$ = 0, 1/2, and
     0.375.}
   \label{fig:dmrg_vs_d}
 \end{figure}

DMRG simulations were performed on cylinders of different aspect
ratios.  We found very small entanglement and finite-size effects due
to the weak coupling between the tetramers.
 Figures \ref{fig:dmrg_vs_d}(a) and (b) shows the convergence of the
 ground state energy with the number of states $d$ for a system of
 size $L_x \times L_y = 16 \times 8$.

 For $m=0$ four significant figures in the ground-state energy are
 achieved with moderate effort $d = 800$ while for $m=1/2$ seven
 significant digits can be obtained with just $d = 200$ states.
  This can be attributed to the weak entanglement in these gaped
  phases.
  Results in the manuscript where obtained with six to seven
  significant figures for a lattice size of $L_x \times L_y = 32
  \times 8$ containing 256 spins.
  Typical runs involved 1000 states for the $m=1/2$ phase and up to
  2400 states in the other cases.

The dependence of the off-diagonal structure factor
$S^{+-}(\pi/2,\pi)$ on the number of DMRG states is shown in Figure
\ref{fig:dmrg_vs_d}(c).  The estimated error with $d=1200$ is in the
third significant digit, while for magnetization $m=1/2$ the results
are fully converged.

\end{document}